\documentclass[10pt]{article}
\usepackage{amssymb}
\usepackage{a4wide}

\newtheorem{theorem}{Theorem}
\newtheorem{definition}{Definition}
\newtheorem{proposition}{Proposition}
\newtheorem{corollary}{Corollary}
\newtheorem{lemma}{Lemma}
\newtheorem{claim}{Claim}
\newtheorem{open}{Open Problem}

\def\squareforqed{\hbox{\rlap{$\sqcap$}$\sqcup$}}
\def\qed{\ifmmode\squareforqed\else{\unskip\nobreak\hfil
\penalty50\hskip1em\null\nobreak\hfil\squareforqed
\parfillskip=0pt\finalhyphendemerits=0\endgraf}\fi}
\newenvironment{proof}
  {\noindent \textbf{Proof:}}
  {\qed}

\newcommand{\code}[1]{\overline{#1}}

\newcommand{\ket}[1]{{|#1\rangle}}
\newcommand{\bra}[1]{{\langle #1|}}
\newcommand{\braket}[2]{{\langle #1|#2\rangle}}
\newcommand{\ketbra}[2]{|#1\rangle\langle#2|}
\newcommand{\length}[1]{\ell(#1)}
\newcommand{\given}{\,\raisebox{-.35ex}{\rule{.5pt}{2ex}}\,}
\renewcommand{\H}{\mathcal{H}}
\newcommand{\Sym}{\mathrm{Sym}}
\newcommand{\Trace}{\mathrm{tr}}
\newcommand{\smfrac}[2]{\mbox{$\frac{#1}{#2}$}}
\newcommand{\fidelity}[2]{{\rm Fidelity}(#1,#2)}

\newcommand{\Entr}{H}
\newcommand{\QC}{\mathit{QC}}

\renewcommand{\dim}{\mathrm{dim}}

\newcommand{\Min}{\mathrm{Min}}
\newcommand{\N}{\mathbb{N}}
\newcommand{\E}{\mathcal{E}}

\begin{document}
\title{\textbf{Quantum Kolmogorov Complexity}\\
~ }
\date{}

\author{{Andr\'e Berthiaume}\thanks{
School of CTI, DePaul University, Chicago, \texttt{berthiaume@cs.depaul.edu}}
\and 
{Wim van Dam}\thanks{C.W.I. Amsterdam; Centre for Quantum Computation, 
University of Oxford, \texttt{wimvdam@qubit.org}}
\and 
{Sophie Laplante}\thanks{L.R.I., Universit\'e Paris Sud, \texttt{Sophie.Laplante@lri.fr}}}

\maketitle
\thispagestyle{empty}
\begin{abstract}
In this paper we give a definition for
quantum Kolmogorov complexity.  In the classical setting, the
Kolmogorov
complexity of a string is the length of the shortest program that can
produce this string as its output.  It is a measure of
the amount of innate randomness (or information) contained in the
string.

We define the quantum Kolmogorov
complexity of a qubit string as the length of the shortest \emph{quantum} 
input to a universal quantum Turing machine
that produces  the initial qubit string with high fidelity.
The definition of Vit\'anyi~\cite{Vitanyi} measures the amount of classical
information, whereas we consider the amount of \emph{quantum}
information in a qubit string.
We argue that our definition
is natural and is an accurate representation of
the amount of quantum information contained in a
quantum state.
\end{abstract}


\section{Introduction} 
\label{intro-section}

In classical computations, the Kolmogorov-Solomonoff-Chaitin
(Kolmogorov, for short) complexity of a finite
string is a measure of its randomness.\cite{Chaitin,Kolmogorov,Solomonoff}  
The Kolmogorov complexity
of $x$ is the length of the shortest program which produces
$x$ as its output.  It can be seen as a lower bound on the
optimal compression that $x$ can undergo, and it is closely
related to Shannon information theory.\cite{IT,Shannon}

Kolmogorov complexity has been shown to have a windfall of
applications in fields as diverse as learning theory,
complexity theory, combinatorics and graph theory, analysis of
algorithms, to name just a few.

With the advent of quantum computation, it is natural to
ask what is a good definition for the Kolmogorov complexity of
quantum strings.  
Our goal in this paper is to argue  that our definition is
a natural and robust measure the amount of quantum information
contained in a quantum string, which has several appealing properties.

Recently, Paul Vit\'anyi~\cite{Vitanyi} has also proposed a 
definition for quantum algorithmic complexity.  Our definition
differs significantly from Vit\'anyi's: the definition he proposes is
a measure of the amount of \emph{classical} information 
necessary to approximate the quantum state.

The paper will be organized as follows: In
Section~\ref{prelim-section},
we give basic notation, definitions, prior work and some
theorems that will be used in proofs in the paper.
In Section~\ref{defn-section} we give our definition of quantum
Kolmogorov complexity.  
In Section~\ref{invariance-section} we prove the invariance theorem.
Section~\ref{properties-section} compares the properties of
quantum and classical Kolmogorov complexity, including
incompressibility,
subadditivity, and the complexity of copies.
Section~\ref{entropy-section}
discusses the relationship with quantum information theory.
We conclude with a discussion of possible extensions and future work.

\section{What is a Good Definition?}

A good definition of quantum Kolmogorov complexity should meet
the following fundamental criteria.  These are intended to
insure that it gives an accurate representation of the
information content of a quantum string.

   \begin{itemize}
   \item It should be robust, that is, invariant under the choice
         of the underlying quantum Turing machine.
    \item It should bear a strong relationship with quantum information
          theory.
    \item It should be closely related to classical complexity
        on classical strings.
   \end{itemize}

However, quantum Kolmogorov complexity
should not be expected to always behave the way classical
Kolmogorov complexity does.
The reader may want to bear in mind
quantum phenomena such as the no-cloning theorem, whose consequences
we will discuss later in the paper.\cite{nocloning}

\subsection{Critical issues}

A first attempt at defining
quantum Kolmogorov complexity of a qubit string $X$ is to
consider
the  length of the shortest quantum program that produces  $X$
as its output.  There are many questions that arise from
this `definition'.

\noindent \textbf{Bits or qubits?}  The first question to consider
is whether we want to measure the amount of algorithmic information
of a string in bits, or in qubits.  Note that bit strings (programs) 
are countable, whereas qubit strings are uncountable, so any definition
that measures in bits would have to overcome this apparent
contradiction.  Paul Vit\'anyi~\cite{Vitanyi} considers classical
descriptions of qubit strings, whereas we consider qubit descriptions.

\noindent \textbf{Exact or inexact?}
What does `produce' mean?  Is a minimal program required to
produce the string $X$ exactly, or only up to some fidelity?
In the latter case, is the fidelity a constant?  Otherwise,
how is it parameterized?  (For exact simulation, we can only
hope to simulate a subclass of the Turing machines, say by
restricting the set of possible amplitudes.  What would be
a reasonable choice?)  We will use an approximation scheme.

\noindent \textbf{What model of computation?}
Size of quantum circuits
is not an appropriate measure since large circuits may be very
simple to describe.  The Turing machine model is
the appropriate one to consider.

\noindent \textbf{What is meant by `quantum program?'}
A program for a quantum Turing machine
is its input, and if we want to count program length
in qubits, we must allow for `programs' to be arbitrary
qubit strings.  (These can be viewed as programs
whose code may include some auxiliary `hard-coded'
qubit strings.)

\noindent \textbf{One-time description or multiple generation?}
In the classical setting, the program that prints the string $x$ can
be run as many times as desired.  Because of the no-cloning theorem
of quantum physics however, we cannot assume that the shortest
program can be run several times to produce several copies
of the same string. This may be due to the fact that it is not
possible to recover the program without losing its output.
There is also a second reason not to choose the multiple generation
option.
The complex-valued parameters $\alpha$ and $\beta$ of a 
qubit $\ket{q}=\alpha\ket{0}+\beta\ket{1}$ can contain an unbounded
amount of information.  If we would be able to reproduce $q$ 
over and over again, then we would have to conclude that 
the single qubit $q$ contains an unlimited amount of information. 
This contradicts the fact that the quantum mechanical system of 
$q$ can only contain one bit of information.\cite{Holevo}
For the above two reason, we will not require a `reusability' condition.

\section{Preliminaries}
\label{prelim-section}

We start with some notation, definitions, and results
that will be used to prove
the results in this paper.

\subsection{Notation}

We use $x$,$y$,\ldots to denote finite, classical Boolean strings.
When we write $\ket{x}$, we mean the quantum state vector in the
standard basis that corresponds to the classical string $x$.
In general we use $\phi,\psi,\ldots$ to denote quantum pure
states. Mixed states are represented by the letters 
$\rho, \sigma$ etc.
We also use uppercase letters $X,Y,\ldots$ for (mixed) 
quantum states that are strings of qubits.
The terms quantum state, qubit string, and quantum register
are used interchangeably (sometimes to emphasize the purpose of 
the quantum state at hand.)  Lower-case letters $i,j,k,l,m,n$
denote integer indices or string lengths.

For classical strings over the alphabet $\Sigma=\{0,1\}$,
$\length{x}$ denotes the length of the string.
For finite sets $A, |A|$ denotes the cardinality of the set.
Concatenation of $x,y$ is written as the juxtaposition $xy$,
and the $n$-fold concatenation of $x$ is written $x^n$.

For Hilbert spaces, we write $\H_d$ for the $d$-dimensional
Hilbert space and $\H^m$ for the $m$-fold tensor product space
$\H\otimes\cdots\otimes \H$. 
A pure quantum state $\phi$ represented as a vector in such 
a Hilbert space is denoted by the ket $\ket{\phi}$. 
The \emph{fidelity} between two pure states $\phi$ and $\psi$ is
the absolute value of the inner product of the two vectors:
$|\braket{\phi}{\psi}|$ (although some authors use the 
square of this value). 

We slightly abuse notation by sometimes letting the state 
symbols $\phi,\rho,\ldots$ also stand for the corresponding 
density matrices. 
Hence, a pure state $\phi$ as a Hilbert space vector is 
denoted by $\ket{\phi}$, whereas its density
matrix $\ketbra{\phi}{\phi}$ can also be denoted by $\phi$.

A density matrix can always be decomposed as a mixture of pure,
orthogonal states: $\rho = \sum_i p_i \ketbra{\phi_i}{\phi_i}$,
with $p_1,p_2,\ldots$ a probability distribution over the
mutually orthogonal states $\phi_1,\phi_2,\ldots$.
The matrix $\rho$ represents a pure state if and only if
$\rho^2=\rho$, in which case we can also say $\sqrt{\rho}=\rho$. 
The square root of a general mixed state is described by
\[ \sqrt{\rho} ~=~  
\sqrt{\sum_i {p_i \ketbra{\phi_i}{\phi_i}}}
~=~ \sum_i {\sqrt{p_i} \ketbra{\phi_i}{\phi_i}}.\]
We use the above rule for the generalization of the fidelity
to mixed states. The fidelity between two density matrices 
$\rho$ and $\sigma$ is defined by
\begin{eqnarray}
\fidelity{\rho}{\sigma} & = & 
\Trace\left({\sqrt{\sqrt{\rho}\cdot \sigma \cdot \sqrt{\rho}}}\right).
\end{eqnarray}
For pure states $\phi$ and $\psi$, the above definition coincides 
again with the familiar $|\braket{\phi}{\psi}|$.
If $\fidelity{\rho}{\sigma} = 1$, then $\rho=\sigma$, and vice versa.

An ensemble $\E$ is specific distribution 
$p_1,p_2,\ldots$ over a set of (mixed) states $\rho_1,\rho_2,\ldots$. 
We denote this by $\E=\{(\rho_i,p_i)\}$. 
The average state of such an ensemble $\E$ is $\rho = \sum_i p_i\rho_i$.
An average state corresponds to several different ensembles.
When an ensemble is used to produce a sequence of states $\rho_i$
according to the probabilities $p_i$, we speak of a 
\emph{source} $\E$.

The length of a quantum state is denoted by $\length{X}$,
by which we mean the smallest $l$ for which
$X$ sits in the $2^l$-dimensional Hilbert space 
(in the standard basis).

A transformation $\$$ on the space of density matrices 
is allowed by the laws of quantum mechanics if and if only it
is a completely positive, trace preserving mapping.

\subsection{Classical Kolmogorov complexity}

The Kolmogorov complexity of a string, in the classical setting, is the
length of the shortest program which prints this string on an empty
input.\cite{LiVitanyi}

Formally, this is stated first relative to a partial computable
function, which as we know can be computed by a Turing machine.

\begin{definition}
Fix a Turing machine $T$ that computes the partial
computable function $\Phi$.
For any pair of strings $x,y \in \{0,1\}^*$, the Kolmogorov
complexity of $x$ relative to $y$ (with respect to $\Phi$) 
is defined as
\begin{eqnarray*}
C_\Phi(x\given y) & =  & {\Min}\{\length{p} : \Phi(p,y) = x\}.
\end{eqnarray*}
\end{definition}
When $y$ is the empty string, we simply write $C_\Phi(x)$.
Also the notation $C_T(x\given y)$ is used.

The key theorem on which rests the robustness of Kolmogorov complexity
is the \emph{invariance theorem}.  This theorem states that the length
of shortest programs does not depend by more than an additive constant
on the
underlying Turing machine.
In the classical case, this theorem is proven with the existence of
a universal Turing machine.  This machine has two inputs: a finite
description of the original Turing machine, and the program that this
Turing machine executes to output the string.

More formally, the invariance theorem in the classical case can be
stated as follows.

\begin{theorem}
There is a universal partial computable function $\Phi_0$
such that for any partial computable $\Phi$
and pair of strings $x, y$,
\begin{eqnarray*}
C_{\Phi_0}(x\given y) & \leq & C_{\Phi}(x\given y)+ c,
\end{eqnarray*}
where $c$ is a constant depending only on $\Phi$.

\end{theorem}

Giving an invariance theorem
will be key to showing that quantum Kolmogorov complexity
is robust.

Since for any string $x$ of length $n$, $C(x) \leq n+O(1)$,
a string which has complexity at least $n$ is called
\emph{incompressible}.  The existence of incompressible
strings is a crucial fact of Kolmogorov complexity.
\begin{proposition}
For every string length $n$, there is a string $x$ of
length $n$ such that $C(x) \geq n$.
\end{proposition}

The proof that there exists incompressible strings is a simple
application of the pigeonhole principle.  By comparing the number
of strings of length $n$ ($2^n$) and the number of programs
of length smaller than $n$ ($2^n-1$ in total), 
one must conclude that there is at
least one string of length $n$ which is not the output of any of the
program of length $<n$.

\subsection{Entropy of classical sources}

The Shannon entropy of a random source that emits symbols from
an alphabet is a measure of the amount of randomness in the
source.\cite{IT,Shannon}
\begin{definition}
Let $A$ be a random source that emits letter $x_i$
(independently) with probability $p_i$.  The Shannon
entropy $\Entr$ of $A$ is $\Entr(A)=-\sum_i {p_i \log p_i }$.
\end{definition}

In the classical setting, Kolmogorov complexity and
Shannon entropy are closely related, as we describe now.
This is an important property of Kolmogorov complexity,
and one would expect a similarly strong relationship
to hold  between quantum Kolmogorov complexity and 
quantum entropy.

Shannon's noiseless coding theorem states that the entropy
corresponds to the average number of bits required to
encode sequences of character emitted by a random source.

\begin{proposition} \textbf{Shannon's noiseless coding \cite{Shannon}:}
Consider a classical channel $A$ that is used to transmit letters
taken from an ensemble $\{(x_i,p_i)\}$, where the $x_i$ are
the letters and $p_i$ their corresponding probabilities. Then
\begin{enumerate}
\item for any $\epsilon,\delta$, there is an $n$ such that
there is an encoding that on $n$ letters encodes
on average the letters with $\Entr(A)+\delta$ bits for which 
the probability
of successfully decoding $P_{\mathrm{success}} \geq 1-\epsilon$;
\item for any $\epsilon,\delta$, there is an $n$ such that
for any $\delta'$, there is an $\epsilon'$ such that
if the channel encodes $n$ letters, each letter with less
than $\Entr(A)-\delta'$ bits per letter, then the probability of
success $P_{\mathrm{success}}\leq 2^{-n(\delta'-\delta)}+\epsilon'$.
\end{enumerate}
\end{proposition}

In the classical case, the Kolmogorov complexity
of a string is bounded by the entropy of a source `likely
to have emitted this string'.  A brief summary of the argument
is included here.  (Details can be found in~\cite[page 180]{LiVitanyi}.)

Let $x$ be a (long) binary string.  It can be broken down into $m$
blocks of
length $k$, where each block is thought of as a character in an
alphabet
of size $2^k$.  Define the frequency $f_i$ of a character
$c_i$ to be the number of times it appears as a block in $x$,
and let $A$ represent the source $\{c_i,f_i/m\}$.
To reconstruct $x$, it suffices to provide the frequency of each
character ($\sum_i \log f_i$ bits)
and then specify $x$ among the strings
that share this frequency pattern.  With some manipulations,
it can be shown that
\begin{proposition}
\label{Kolm-entropy-prop}
\begin{eqnarray*}
C(x) & < & m(\Entr(A)+\gamma),
\end{eqnarray*}
where $A$ is the source defined in the discussion above, 
and $\gamma$ vanishes as $m$ goes to infinity.
\end{proposition}

\subsection{Quantum information theory}

We have seen that in the classical setting, Kolmogorov 
complexity is very closely related to Shannon entropy.
In this section we describe the quantum, or Von Neumann,
entropy, related measures, and important properties which
will be used in the proofs of our results.

\begin{definition}
\textbf{Von Neumann entropy:}
The Von Neumann entropy of a mixed state $\rho$ is 
defined as $S(\rho) = \Trace(-\rho \log \rho)$.
If we decompose $\rho$ into its mutually orthogonal 
eigenstates $\phi_i$, we see that
\begin{eqnarray*}
S(\rho)~ = ~ S\left({\sum_i {p_i \ketbra{\phi_i}{\phi_i}}}\right)~ 
= ~ H(p),
\end{eqnarray*}
where $H(p)$ is the Shannon entropy of the
probability distribution $p_1,p_2,\ldots$
\end{definition}

A source $\E=\{(\rho_i,p_i)\}$ has an 
associated Von Neumann entropy $S(\rho)$ of the
average state $\rho=\sum_i p_i\rho_i$.
Schumacher's noiseless coding theorem~\cite{Schumacher} shows
how to obtain an encoding with average letter-length $S(\rho)$ 
for a source of pure states, where the fidelity of the encoding 
goes to $1$ as the number of letters emitted by the source 
goes to infinity.  (A survey can be found in Preskill's lecture 
notes~\cite[page 190]{Preskill} or in Nielsen's 
thesis~\cite[Chapter~7]{Nielsen}.)

We will use a slightly stronger result, which gives a
universal compression scheme. That is, one that does
not depend on the source itself, but only on its entropy.  
This result is due to Jozsa et al.~\cite{JH3}, 
building upon the work of Jozsa and Schumacher~\cite{JS}.

\begin{theorem} \textbf{Universal quantum compression~(see \cite{JS,JH3}):}
\label{JH3-coding}
Consider pure state sources $\E=\{(\phi_i,p_i)\}$.
For any $\epsilon, \delta$, there is an 
$n=n(\epsilon,\delta)$ such that for any 
entropy bound $S$, there is an encoding scheme that works for any source 
of Von Neumann entropy at most $S$ that has the following properties.  
Let $\rho = \sum_i p_i \ketbra{\phi_i}{\phi_i}$ be the average state, 
with all $\ket{\phi_i} \in \H_d$,
and $\rho$ has entropy $S(\rho)\leq S$, then
\begin{enumerate}
\item Each $\ket{\phi_i}$ can be encoded by a code word $\sigma_i$, which
    has length $\leq S + \delta + \smfrac{1}{n}(d^2 \log(n+1))$.
\item  For each $i$, $\fidelity{\phi_i}{\sigma_i} \geq 1-\epsilon$.
\end{enumerate}
\end{theorem}

We continue the section by  defining the `$\chi$ quantity' for ensembles.
\begin{definition}
\textbf{Holevo's chi quantity \cite{Holevo}:}
For an ensemble $\E=\{(\rho_i,p_i)\}$, 
with $\rho = \sum_i p_i \rho_i$, 
Holevo's chi quantity equals 
\begin{eqnarray*}
\chi(\E) & = & S(\rho) - \sum_i {p_i S(\rho_i)}.
\end{eqnarray*}
\end{definition}
Note that the $\chi$ quantity depends not only on $\rho$, but also 
on the specific pairs $(p_i,\rho_i)$.

The following monotonicity property of Lindblad and Uhlmann will
be very useful later in the paper.
\begin{theorem}\textbf{Lindblad-Uhlmann monotonicity 
\cite{Lindblad,Uhlmann}:}
\label{monotonicity-theorem}
Let $\E =\{(\rho_i,p_i)\}$ be an ensemble, and $\$$ a completely positive,
trace preserving mapping.
For every such $\E$ and $\$$, it holds that: $\chi(\$(\E)) \leq \chi(\E)$,
where $\$(\E)$ is the transformed ensemble $\{(\$(\rho_i),p_i)\}$.
\end{theorem}

The entropy of finite systems is robust against small
changes.  This continuity of $S$ over the
space of finite dimensional density matrices $\rho$ is also 
called \emph{insensitivity}, and is expressed by the following lemma.
\begin{lemma}\textbf{Insensitivity of Von Neumann entropy 
(see Section II.A in \cite{Wehrl}):}\label{insensitivity}
If a sequence $\rho_1,\rho_2,\ldots$, has
$\lim_{k\rightarrow\infty}{\rho_k}=\rho$, then
also $\lim_{k\rightarrow\infty}{S(\rho_k)}=S(\rho)$.
\end{lemma}
\begin{proof}
The convergence of $\rho_1,\rho_2,\ldots$ to $\rho$ is
understood to use some kind of norm for the density
matrices that is continuous in the matrix entries 
$\bra{i}\rho\ket{j}$.
(The operator norm $|\rho|=\Trace(\rho\rho^*)$, for example.)
The entropy $S(\rho)$ is a continuous function 
of the finite set of eigenvalues of $\rho$. These eigenvalues
are also continuous in the entries of $\rho$.
\end{proof}

Further background on these measures of quantum
information and their properties can be found in 
\cite[Chapter~5]{Preskill}.   Another good source is
Nielsen's thesis~\cite{Nielsen}.

\subsection{Symmetric spaces}

We use the symmetric subspace of the Hilbert space to show some
of our results on copies of quantum states. 
Let $\H_D$ be a Hilbert space of dimension $D$
with the basis states labeled $\ket{1},\ldots,\ket{D}$.
The symmetric subspace $\Sym({\H_D^{m}})$ of the $m$-fold tensor
product
space $\H_D^{m}$ is a subspace spanned by as many basis vectors
as there
are multisets of size $m$ of $\{1,\ldots,D\}$.
Let $A=\{i_1,\ldots,i_m\}$ be such a multiset of $\{1,\ldots,D\}$.
Then, $\ket{s_A}$ is the normalized superposition of all the
different permutations of $i_1,\ldots,i_m$.
The set of the different vectors $\ket{s_A}$ 
(ranging over the multisets  $A$)
is an orthogonal basis of the symmetric subspace 
$\Sym(\H^m_D)$.
Hence the dimension of the 
symmetric subspace is $m+D-1 \choose D-1$.
(This is because choosing a multiset is the same thing as
splitting $m$ consecutive elements into $D$ (possibly empty)
intervals, where the size of $i$th interval represents the number 
of times the $i$th element appears in the multiset.  The number
of ways of splitting an interval of size $m$ into $D$
intervals is $m+D-1 \choose D-1$.)

An equivalent definition of the symmetric subspace is that
it is the smallest subspace that contains all the states of the
form $\ket{\phi}^m$, for all $\ket{\phi}\in \H_D$.  
(For more on the symmetric subspace and its properties, see the paper by 
Barenco et al.~\cite{symmetrisation}.)

\subsection{Accumulation of errors}

The following lemma is used to bound the error introduced when
composing two inexact quantum procedures.

\begin{lemma} \textbf{Fidelity of composition:}
\label{accumulation-lemma}
If $\fidelity{\rho}{\rho'}\geq 1-\delta_1$ and 
$\fidelity{\rho'}{\rho''}\geq 1-\delta_2$, then
$\fidelity{\rho}{\rho''}\geq 1-2\delta_1-2\delta_2 $.
\end{lemma}
\begin{proof}
This follows from the fact that the fidelity between two
mixed states $\rho$ and $\sigma$ equals the maximum 
`pure state fidelity' $|\braket{\phi}{\psi}|$,
where $\phi$ and $\psi$ are `purifications' of
$\rho$ and $\sigma$. 
(See \cite{Fuchs} for more details on this.)
\end{proof}

In order to give bounds on the complexity of
several copies of a state, as we do in Section~\ref{copies-section},
we need the following bound on the total error in the
$n$-fold tensor product of the approximation of a given state.

\begin{lemma}
\label{fidelity-of-copies-lemma}
Let $\rho^n$ and $\sigma^n$ be the $n$-fold 
copies of the mixed states $\rho$ and $\sigma$, then 
$\fidelity{\rho^n}{\sigma^n} =  (\fidelity{\rho}{\sigma})^n$.
\end{lemma}

\begin{proof}
This follows directly from the definition 
$\fidelity{\rho}{\sigma} = 
\Trace\left({\sqrt{\sqrt{\rho} \cdot \sigma \cdot \sqrt{\rho}}}\right)$.
\end{proof}

\section{Quantum Kolmogorov Complexity}
\label{defn-section}

We define the \emph{quantum Kolmogorov complexity} $\QC$ of a string
of qubits, relative to a quantum Turing machine $M$, as the length of
the shortest qubit string which when given as input to $M$, produces
on its output register the qubit string. (Note that we only allow $M$
that have computable transition amplitudes.
See the articles \cite{BV,Deutsch}, and particularly Definition~3.2.2
in \cite{BV}, for a further description of this 
computational model.)

\subsection{Input/Output Conventions}

We give some precisions about what is meant by `input' and `output'.

We consider quantum Turing machines with two heads on two 
one-way infinite tapes. 
We allow the input tape to be changed.
This is required: for example, the contents of the input may 
have to be moved to the output tape.

For a QTM $M$ with a single input, when we say
$M$ starts with input $Y$, we mean that M starts
with the quantum state $\ket{Y\$00\cdots}$ on its input tape,
and $\ket{00\cdots}$ on the output tape.  The $\$$ symbol
is a special endmarker (or blank) symbol.

Note that testing for the end of the
input can be done without disturbing the input, since
we assume that the `\$' state is orthogonal to the `0' and `1'
states.   (This is analogous to the classical case,
where where Turing machine inputs are encoded in a
three-letter alphabet; nevertheless we consider
the actual input to be encoded only over the characters
0 and 1.)

A string is a proper input if the endmarker symbol appears
only once and is not in superposition with any other
position of the tape.  We dismiss any non-proper inputs.

For a QTM with multiple inputs, we also assume that
there is a convention for encoding the multiple inputs so that
they can be individually recovered.  For example,
when we write $M(P,Y)$, we may assume that the input
tape is initialized to $\ket{1^{\length{P}}PY\$00\cdots}$.
We only count the length of $X$ and $Y$ for the length 
of the input.
Likewise, for multiple outputs, if we write
$M(P,Y)=(X_1,X_2)$, we mean that $X_1$ and $X_2$ must be
encoded according to a prearranged convention
so that $X_1$ and $X_2$ can be recovered individually
from the output tape.

(Note that we do not define prefix-free complexity in
this paper.  The programs themselves need not be prefix-free.)

We let $M^T(X)$ denote the contents of the output tape
after $T$ steps of computation.  We consider only QTMs
which do not modify their output tape after they have 
halted.  (Because of reversibility, they may modify
the input tape after reaching the halting state.)
The output $M(X)$ is the content of the output
tape at any time after $M$ has stopped changing
its output tape.

\subsection{Definitions}

For some fidelity function $f:\N\rightarrow [0,1]$ we will
now define the corresponding quantum Kolmogorov complexity.

\begin{definition}\textbf{Quantum Kolmogorov complexity with 
fidelity \textit{f}:}
For any quantum Turing machine $M$ and qubit string $X$,  
the $f$-approximation quantum Kolmogorov
complexity, denoted $\QC^{f}_M(X)$,
is the length of the smallest qubit string $P$ 
such that for any fidelity parameter $k$ we have
$\fidelity{X}{M(P,1^k)}\geq f(k)$.
\end{definition}

Note that we require that the same string $P$ be used for all
approximation parameters $k$.

We will say that program $P$ $M$-computes $X$ with fidelity $f(k)$
if $\forall k,\; \fidelity{M(P,1^k)}{X} \geq f(k)$.

If $f$ is the constant function $1$, we have the following 
definition.

\begin{definition} \textbf{Quantum Kolmogorov complexity with 
perfect fidelity:}\label{qc_perfect}
The perfect fidelity quantum Kolmogorov complexity is $\QC^1_M(X)$.
\end{definition}

The problem with this definition is that it is not known whether an
invariance theorem can be given for the ideal Kolmogorov complexity.
This is because the invariance theorems that are known for quantum
computers deal with \emph{approximating} procedures.
We therefore prove an invariance theorem
for a weaker, limiting version, where the output of $M$ must have
high fidelity with respect to the target string $X$: 
$\fidelity{X}{M(P)}\approx 1$.

\begin{definition}
\textbf{Quantum Kolmogorov complexity with bounded fidelity:}
For any constant $\epsilon < 1$, $\QC^\epsilon_M(X)$
is the constant-fidelity quantum Kolmogorov
complexity.
\end{definition}

There are two problems with this definition.  First,
it may be the case that some strings are very easy to
describe up to a given constant, but inherently very hard to 
describe for a smaller error.  Second, it may be the case
that some strings are easier to describe up to a given
constant on one machine, but not on another machine.
For these two reasons, this definition does not appear
to be robust.

A stronger notion of approximability is the existence of an
approximation \emph{scheme.}  (See, for example, the book by
Garey and Johnson~\cite[Chapter~6]{GareyJohnson} for more on
approximation algorithms and approximation schemes.)

For constant-approximability, different algorithms (with different
sizes) can exist for different constants.  
In an approximation scheme, a single
program takes as auxiliary input an approximation parameter $k$,
and produces an output that approximates the value we want
within the approximation parameter.  This is the model
we wish to adopt for quantum Kolmogorov complexity.

\begin{definition}\label{qc_converging}
\textbf{Quantum Kolmogorov complexity with fidelity converging to 1:}
The complexity $\QC^{\uparrow 1}_M(X)$ 
is equal to $\QC^f_M(X)$, where $f(k)=1-\frac{1}{k}$.
\end{definition}

We choose to encode the fidelity parameter in unary, and
the convergence function to be $f(k)=1-\frac{1}{k}$ so
that the model remains robust when polynomial time bounds
are added.  We discuss this further in Section~\ref{invariance-section}.

We may also define $\QC^{\uparrow 1}_M(X\given Y)$, the complexity of
producing $X$ when $Y$ is given as an auxiliary input,
in the usual way.

\section{Invariance}
\label{invariance-section}

To show that our definition is robust we must show that the complexity
of a qubit string does not depend on the underlying quantum Turing machine.

We use the following result, proved in the paper of Bernstein and
Vazirani \cite{BV}.  To be precise, we use the notation
$\code{M}$ to denote the classical description of the
quantum Turing machine $M$.  (Recall that we only consider
quantum Turing machines whose amplitudes can be computed to
arbitrary precision with a finite classical 
description.)

\begin{theorem}
\textbf{Universal quantum Turing machine (see \cite{BV}):}
There exists a universal quantum Turing machine $U$
that has a finite classical description such that the following
holds.  For any quantum Turing machine $M$ (which has
a finite classical description), for any pure state
$X$, for any approximation parameter $k$, and any number of time steps $T$, 
$\fidelity{U(\code{M},X,1^k,T)}{M^T(X)} \geq 1-\frac{1}{k}$.
Recall that $M^T$ is the contents of the output tape of $M$ after $T$ time
steps.
\end{theorem}

\begin{theorem}
There is a universal quantum Turing machine  $U$
such that for any quantum Turing machine $M$
and qubit strings $X$,
\begin{eqnarray*}
 \QC^{\uparrow 1}_{U}(X) & \leq &  \QC^{\uparrow 1}_{M}(X)+ c_M,
\end{eqnarray*}
where $c_M$ is a constant depending only on $M$.
\end{theorem}
\begin{proof} 
The proof follows from the existence of a universal quantum
Turing machine, as proven by Bernstein and Vazirani~\cite{BV}.
Let $U$ be this UTM as mentioned above.
The constant $c_M$ represents the size  of the finite description
that $U$ requires to calculate the transition amplitudes of the machine $M$.
Let $P$ be the state that
witness that $\QC_M^{\uparrow 1}(X)=\length{P}$, and hence 
$\fidelity{X}{M(P,1^{k})} \geq 1-\smfrac{1}{k}$ for every $k$.

With the description corresponding to $c_M$, $U$ can simulate
with arbitrary accuracy the behavior of $M$. Specifically, 
$U$ can simulate machine $M$ on input $(P,1^{4k})$ with a fidelity 
of $1-\smfrac{1}{4k}$. 
Therefore, by Lemma~\ref{accumulation-lemma}, 
$\fidelity{X}{U(M,P,1^{4k})} \geq 1-\smfrac{1}{k}$.
\end{proof}

The same holds true for the conditional complexity, that is,
$\exists U \forall M, X,Y$, $\QC^{\uparrow 1}_{U}(X\given Y)
\leq \QC^{\uparrow 1}_{M}(X\given Y)+ c_M$.

Henceforth, we will fix a universal quantum Turing machine
$U$ and simply write  $\QC(X)$ instead of $\QC^{\uparrow 1}_{U}(X)$.
Likewise we write  $\QC(X|Y)$ instead of $\QC^{\uparrow 1}_{U}(X|Y)$.
We also abuse notation and write $M$ instead of $\code{M}$
to represent the code of the quantum Turing machine $M$ used as an
input to the universal Turing machine.

We may also define time-bounded
$\QC$ is the usual way, that is, fix $T:\N\rightarrow \N$
a fully-time-computable function.  Then $\QC^T(X|Y)$ is the
length of the shortest program which on input $Y,1^k$,
produces $X$ on its output tape after $T(\length{X}+\length{Y})$
computation steps.  The Bernstein and Vazirani
simulation entails a polynomial time blowup
(polynomial in the length of the input and the length of
the fidelity parameter encoded in unary), so there is 
a polynomial time blowup in the corresponding invariance theorem.

The simplest application of the invariance theorem is the
following proposition.

\begin{proposition}
For any qubit string $X$, $\QC(X)\leq \length{X} + c$,
where $c$ is a constant depending only on our choice of the
underlying universal Turing machine.
\end{proposition}
\begin{proof}
Consider the quantum Turing machine $M$ that moves its input to the
output tape, yielding $\QC_M(X) = \length{X}$.
The proposition follows by invariance.
\end{proof}
\section{Properties of Quantum Kolmogorov Complexity}
\label{properties-section}

In this section we compare classical and quantum Kolmogorov
complexity by examining several properties of both.  We
find that many of the properties of the classical complexity,
or natural analogues thereof, also hold for the quantum complexity.
A notable exception is the complexity of $m$-fold copies of
arbitrary qubit strings.

\subsection{Correspondence for classical strings}

We would like to show that for classical states, classical and 
quantum Kolmogorov complexity coincide, up to a constant additive term.

\begin{proposition}
\label{correspondence-theorem}
For any finite, classical string $x$,
$\QC(x) \leq C(x)+ O(1)$.
\end{proposition}
(The constant hidden by the big-$O$ notation depends only on the
underlying universal Turing machine.)

\begin{proof}
This is clear: the universal quantum computer can also simulate
any classical Turing machine.
\end{proof}

We leave as a tantalizing open question whether the
converse is also true, that is:
\begin{open}
Is there a constant $c$ such that 
for every finite, classical string $x$,
$C(x) \leq \QC(x) + c$?
\end{open}

\subsection{Quantum incompressibility}

\label{incompressibility-section}


In this section, we show that there exist quantum-incompressible
strings.

Our main theorem is a very general form of the incompressibility theorem.
We state some useful special cases as corollaries. 

Assume we want to consider the minimal-length programs that
describe a set of quantum states.  In general, these may be pure 
or mixed states.  We will use the following notation
throughout the proof.  The mixed states $\rho_1,\ldots,\rho_M$ be 
are the target strings (those we want to produce as output). 
Their minimal-length programs 
will be  $\sigma_1,\ldots, \sigma_M$, respectively. 
The central idea is that if the states $\rho_i$ are
sufficiently different, then the programs $\sigma_i$
must be different as well. We turn this into a 
quantitative statement with the use of the insensitive 
chi quantity in combination with the monotonicity of 
quantum mechanics.
 
\begin{theorem}\label{thm:incompres}
For any set of strings $\rho_1,\ldots,\rho_M$ such that 
$\forall i, QC(\rho_i)\leq l$, this $l$ is bounded from below by
\begin{eqnarray*}  
 l & \geq &
 S(\rho) -  \smfrac{1}{M}\sum_i S(\rho_i),
\end{eqnarray*}
where $\rho$ is the `average' density matrix
$\rho = \smfrac{1}{M}\sum_i \rho_i$.

(Stated slightly differently, this says that there is an $i$
such that $\QC(\rho_i)\geq S(\rho)-\smfrac{1}{M}\sum_i S(\rho_i)$.)
\end{theorem}
\begin{proof}
Take $\rho_1,\ldots,\rho_M$ and their minimal programs
$\sigma_1,\ldots,\sigma_M$ (and hence 
$QC(\rho_i) = \length{\sigma_i}$). 
Let $\$^k$ be the completely positive, trace preserving map
corresponding to the universal QTM $U$ with fidelity parameter $k$. 
With this, we define the following three uniform ensembles:
\begin{itemize}
\item{the ensemble $\E=\{(\rho_i,\smfrac{1}{M}\}$ of the original strings,}
\item{$\E_\sigma$ the ensemble of programs 
$\{(\sigma_i,\smfrac{1}{M})\}$, and}
\item{the ensemble of the $k$-approximations $\tilde{\E}^k = \$^k(\E_\sigma) = 
\{(\tilde{\rho}^k_i,\smfrac{1}{M})\}$, with $\tilde{\rho}_i^k = \$^k(\sigma_i)$.}
\end{itemize}
By the monotonicity of Theorem~\ref{monotonicity-theorem} we know
that for every $k$, $\chi(\tilde{\E}^k)\leq \chi(\E_\sigma)$.
The chi factor of the ensemble $\E_\sigma$ is upper bounded
by the maximum size of its strings: 
$\chi(\E_\sigma)\leq \max_i\{\length{\sigma_i}\} \leq l$.
Thus the only thing that remains to be proven is that $\chi(\tilde{\E}^k)$,
for sufficiently big $k$, is `close' to $\chi(\E)$. This will be done 
by using the insensitivity of the Von Neumann entropy.

By definition, for all $i$,
$\lim_{k\rightarrow \infty}{\fidelity{\rho_i}{\tilde{\rho}^k_i}} = 1$,
and hence $\lim_{k\rightarrow \infty}{\tilde{\rho}^k_i} = \rho_i$.
Because the ensembles $\E$ and $\tilde{\E}^k$ have only a finite number 
($M$) of states, we can use Lemma~\ref{insensitivity}, to obtain 
$\lim_{k \rightarrow \infty}{\chi(\tilde{\E}^k)} = \chi(\E)$.
This shows that for any $\delta>0$, there exists a $k$ such that 
$\chi(\E)-\delta \leq \chi(\tilde{\E}^k)$. With the above inequalities
we can therefore conclude that $\chi(\E)-\delta \leq l$ holds for 
arbitrary small $\delta>0$, and hence that $l\geq \chi(\E)$.
\end{proof}

The following four corollaries are straightforward with the
above theorem.

\begin{corollary}
For every length $n$, there is an incompressible
classical string of length $n$.
\end{corollary}

\begin{proof}
Apply Theorem~\ref{thm:incompres} to the set of classical strings
of $n$ bits: $\rho_x = \ketbra{x}{x}$ for all $x\in\{0,1\}^n$.
All $\rho_x$ are pure states with zero Von Neumann entropy, hence
the lower bound on $l$ reads $l \geq S(\rho)$.
The average state $\rho = 2^{-n}\sum_{x}\ketbra{x}{x}$ is the 
total mixture $2^{-n}I$ with entropy $S(\rho)=n$, 
hence indeed $l \geq n$.
\end{proof}

\begin{corollary}
\label{log-size-cor}
For any set of orthogonal pure states $\ket{\phi_1},$ $\ldots,$ $\ket{\phi_M}$, 
the smallest $l$ such that for all $i$, $\QC(\phi_i) \leq l$
is at least $\log M$.  (Stated differently, there is an $i$ such
that  $\QC(\phi_i) \geq \log M$.)
\end{corollary}

\begin{proof}
All the pure states have zero entropy $S(\phi_i)=0$,
hence by Theorem~\ref{thm:incompres}: $l \geq S(\rho)$.
Because all $\phi_i$s are mutually orthogonal, this 
Von Neumann entropy $S(\rho)$ of the average state 
$\rho = \smfrac{1}{M}\sum_i {\ketbra{\phi_i}{\phi_i}}$
equals $\log M$.
\end{proof}

\begin{corollary}
For every length $n$, at least $2^n-2^{n-c}+1$ qubit strings 
of length $n$ have complexity at least $n-c$.
\end{corollary}

\begin{corollary}\label{incompres-pure}
For any set of pure states $\ket{\phi_1},\ldots, \ket{\phi_M}$,
the smallest $l$ such that for all $i$, $\QC(\phi_i) \leq l$
is at least $S(\rho)$, where 
$\rho=\smfrac{1}{M}\sum_i \ketbra{\phi_i}{\phi_i}$.
\end{corollary}

\subsection{The complexity of copies}
\label{copies-section}

A case where quantum Kolmogorov complexity behaves differently from
classical Kolmogorov complexity is that, in general, the relation
$C(x^m)\leq C(x) + O(\log m)$ does not hold, as we show below.
We give an upper and a lower bound for the Kolmogorov complexity of
$X^{m}$.

\begin{theorem}
$\QC(X^m) \leq \log{{m+2^{\QC(X)}-1}\choose{2^{\QC(X)}-1}} + O(\log m)
+ O(\log \QC(X)).$
\end{theorem}
\begin{proof}
First we sketch the proof, omitting the effect of the approximation.
Consider any qubit string $X$ whose minimal-length program
is $P_X$.  To produce $m$ copies of $X$, it suffices to
produce $m$ copies of $P_X$ and make $m$ runs of $P_X$.

Let $l$ be the length of $P_X$; we call $\H$ the $2^l$-dimensional 
Hilbert space.  Consider
$\H^{m}=\H \otimes \cdots \otimes \H$, the $m$-fold
tensor product of $\H$.
The symmetric subspace $\Sym({\H^{m}})$ is
$d$-dimensional, where $d= {m+2^l-1\choose 2^l-1}$.
The state $P_X^m$ sits in this symmetric subspace,
and can therefore be encoded exactly using 
$\log d + O(\log m)+O(\log l)$ qubits, where the $O(\log m)$ 
and $O(\log l)$ terms are used to describe the rotation onto $\Sym(\H^m)$.
Hence, the quantum Kolmogorov complexity of $X^m$ is bounded 
from above by $\log d + O(\log m)+O(\log l)$ qubits.

For the full proof, we will need to take into account the 
effect of the imperfect fidelities of the different computations.

To achieve a fidelity of $1-\smfrac{1}{k}$, we will compute $m$
copies of the minimal program $P_X$ to a fidelity of $1-\smfrac{1}{4km}$.
On each copy, we simulate the program with fidelity
of $1-\smfrac{1}{4km}$, and thus obtain the strings $\tilde{X}_i$ ($1\leq i\leq m$),
each of which has
(according to Lemma~\ref{accumulation-lemma}) fidelity $1-\smfrac{1}{km}$ with
the target string $X$.  By 
Lemma~\ref{fidelity-of-copies-lemma}
we get a total fidelity  of at least $1-\smfrac{1}{k}$.

We now proceed to the details of the proof.  First
we introduce some notation.

Assume that for some QTM $M$, $QC_M(X) \leq \length{P_X}=l$, 
where $P_X$ $M$-computes $X$
(with fidelity $1-\frac{1}{k}$ for any $k$.)

Let $R$ be the rotation that takes qubit strings 
$X^m\in\Sym({\H^{m}})$ to qubit strings of length 
$\lceil \log(\dim(\Sym({\H^{m}})))\rceil$.  More precisely,
$R$ is the rotation that takes the $i$th basis state of 
$\Sym({\H^{m}})$ to the $i$th classical basis state of the
Hilbert space of dimension $2^{\lceil \log(\dim(\Sym({\H^{m}})))\rceil}$.

For any fidelity parameter $\delta$,
$R^{-1}$ can be computed efficiently and to arbitrary
precision.  By that we mean that for any $\delta$, there
is a transformation $R^{-1}_{\delta}$ for which the following holds:
Let $Z=R(X^m)$ for some $X\in \H$.  If 
$\widetilde{X^m}=R^{-1}_{\delta}(Z)$, then for each $i$,
the mixed state $\tilde{X}_i$
obtained from $X$ by tracing out all components that do not
correspond to the $i$th copy of $X$, is such that
$\fidelity{X}{\tilde{X}_i}\geq 1-\delta$.

We now define the program that witnesses the upper bound on
$\QC(X^m)$ claimed in the theorem.

Let $M'$ be the quantum Turing machine that does the following
on input $(Z,l,m,1^k)$.
\begin{enumerate}
\item Computes $Z'=R^{-1}_{1/4km}(Z)$.  (When $Z$ is an $m$-proper input, 
   which we specify below, then $Z'\approx Y^m$ for some $Y \in \H$.)
\item On each `copy' $\tilde{Y}_i$ of $Y$, runs the QTM 
      $M(\tilde{Y}_i,1^{4km})$.  (That is, $\tilde{Y}_i$ is the result of
      tracing out all but the positions of $Z'$ that correspond
      to the $i$th block of $l$ qubits.)
\end{enumerate}

The input $Z$ is an `$m$-proper input' if for some $Y$, 
$Z=R(Y^m)$.  (Note that $Z$ is exactly $R(Y^m)$, not
an approximation up to some fidelity.)

If we run the above QTM $M'$ on input $(R(P_X^m),l,m,1^k)$
then the output of this $M'$ is $M'(R(P_x^m),l,m,1^k)
 =\widetilde{X^m}=\tilde{X}_1\cdots\tilde{X}_m$.
(Recall that $l$ is the length of $P_X$.)

It remains to show the following claims.
\begin{claim} \label{copies-claim1}
$\fidelity{\widetilde{X^m}}{X^m} \geq 1- \frac{1}{k}$.
\end{claim}

\begin{claim}
\label{copies-claim2} The length of the program above for $M'$ is 
$\leq \log d_{l,m} + O(\log l) + O(\log m)$, where $d_{l,m} = {m+2^l-1\choose 2^l-1}$.
\end{claim}

Claim~\ref{copies-claim2} follows immediately from the fact
that the total length of the inputs 
$R(P_X^m),l,m$ is $\log d + O(\log l) + O(\log m)$.

We prove Claim~\ref{copies-claim1}.
Since we chose a precision $\delta=\smfrac{1}{4km}$ in step~1,
$\forall i$,  $\fidelity{P_X}{\tilde{Y}_i} \geq 1-\smfrac{1}{4km}$.
Furthermore, since the computation at step~2 introduces at
most an error of $\smfrac{1}{4km}$,
$\forall i$, $\fidelity{X}{\tilde{X}_i} \geq 1-\smfrac{1}{km}$  
(by Lemma~\ref{accumulation-lemma}.)  Therefore
by Lemma~\ref{fidelity-of-copies-lemma}, 
$\fidelity{\widetilde{X^m}}{X^m}\geq (1-\smfrac{1}{km})^m \geq 
1-\smfrac{1}{k}$. 
This completes the proof of Claim~\ref{copies-claim1}.

Claim~\ref{copies-claim1} and Claim~\ref{copies-claim2} together
give us that $QC_{M'}(X^m)\leq \log d_{l,m} + O(\log l) + O(\log m)
 \leq  \log d_{n,m} + O(\log n) + O(\log m)$,
where $n$ is the length of $X$ and an upper bound on its
complexity.  By invariance, we can conclude that
$QC(X^m)\leq \log d_{n,m} + O(\log n) + O(\log m)+ O(1)$, which 
proves the theorem.
\end{proof}

This upper bound is also very close to being tight for some $X$, 
as we show in the next theorem.

\begin{theorem}
\label{copies-theorem}
For every $m$ and $n$, there is an $n$-qubit state $X$ such that
$\QC(X^m) \geq \log{{m+2^{n}-1}\choose{2^{n}-1}}$.
\end{theorem}

\begin{proof} 
Fix $m$ and $n$ and let $\H$ be the $2^n$-dimensional Hilbert space.
Consider the (continuous) ensemble of all $m$-fold
tensor product states $X^m$: $\E=\{(X^m,\mu)\}$,
where $\mu^{-1}=\int_{X\in \H}{dX}$ is the appropriate normalization 
factor. The corresponding average state is calculated by the 
integral 
$\rho = \mu \int_{X \in \H}{X^m dX}$.
This mixture is the totally mixed state in the symmetric subspace 
$\Sym(\H^m)$ (see Section~3 in \cite{Werner}),
and hence has entropy
$S(\rho) = \log{{m+2^n-1}\choose{2^n-1}}$. 
Because all $X^m$ are pure states, we can use 
Corollary~\ref{incompres-pure}
to prove the existence of a $X$ for which 
$QC(X^m) \geq \log{{m+2^n-1}\choose{2^n-1}}$.
\end{proof}

\subsection{Subadditivity}

Consider the following subadditivity property
of classical Kolmogorov complexity.

\begin{proposition}
\label{subadditivity}
For any $x$ and $y$, $C(x,y)\leq C(x) + C(y\given x) + O(1)$.
\end{proposition}
In the classical case, we can produce $x$, and then
produce $y$ from $x$, and print out the combination of $x$ and $y$.
In the quantum case, producing $Y$ from $X$ may destroy $X$.
In particular, with $X=Y$, the immediate quantum analogue
of Proposition~\ref{subadditivity} would contradict
Theorem~\ref{copies-theorem} (for $m=2$).

A natural quantum extension of this result is as follows.

\begin{proposition}
\label{subadditivity-q}
For any $X,Y$, $\QC(X,Y)\leq \QC(X,X) + \QC(Y\given X) + O(1)$.
\end{proposition}

\section{Quantum Information Theory}
\label{entropy-section}

In this section we establish a relationship between
quantum compression theory and the 
bounded-fidelity version of quantum 
Kolmogorov complexity.

One would like to give a direct analogue of 
Proposition~\ref{Kolm-entropy-prop}.
However, we believe that such a statement does not hold
for quantum Kolmogorov complexity.  The argument can
be summarized as follows.  In the classical case,
given a string $x$, we can define a source $A$ such that
$x$ is in the so-called `typical subspace' of $A$.
This allows us to give a short, exact description of $x$.

In the quantum case, we may also define a quantum source
likely to have emitted a given qubit string $X$ (in an appropriate
tensor space).  However, we do not get that $X$ is \emph{in} the
typical subspace of this source, only that it is \emph{close} to 
the typical subspace.  How close it can be guaranteed to be depends
on the length of $X$.  Therefore, for a fixed string length
$n$, we may not be able to get an encoding of arbitrary high fidelity.

We now prove a slightly weaker statement, for bounded-fidelity
complexity.

\begin{theorem}
Let $U$ be the universal quantum Turing machine from~\cite{BV}.  
Then for any $\epsilon, \delta$ there is an $n$ such that for
any $d$-dimensional $\H$, 
and any qubit string $X=\ket{\phi_1}\otimes\cdots\otimes\ket{\phi_n} \in \H^n,$
\begin{eqnarray*}   
\QC^{\epsilon}_{U}(X) 
  & \leq & n(S(\rho)+\delta+ \smfrac{1}{n}(d^2 \log(n+1))),
\end{eqnarray*}
where $\rho=\smfrac{1}{n}\sum_i \ketbra{\phi_i}{\phi_i}$.
\end{theorem}

\begin{proof}  
Fix $\epsilon, \delta$.  Apply Theorem~\ref{JH3-coding}
with  $\epsilon'=\smfrac{\epsilon}{4},\delta'=\delta,$ and let
$n=n(\epsilon',\delta')$ be the value from the theorem.
Let $\ket{\phi_1}\otimes\cdots\otimes\ket{\phi_n} \in \H^n$ 
be the string for whose quantum Kolmogorov complexity we want
to give an upper bound.  
By  Theorem~\ref{JH3-coding}, item~1,
we get that the length of the
encoding is what was given in the statement of the theorem.
By simulating the decoding algorithm to a precision
of $\smfrac{\epsilon}{4}$, together with Theorem~\ref{JH3-coding}, item~2, 
and Lemma~\ref{accumulation-lemma},
we have that the fidelity of the encoding is at least $1-\epsilon$.
That completes the proof.
\end{proof}

\section{Extensions and Future Work}

We have argued that the $\QC$ of Definition~\ref{qc_converging} 
is a robust notion of Kolmogorov complexity for the quantum setting.  
Nevertheless, it would be interesting
to see if an invariance theorem can be shown for the ideal quantum
Kolmogorov complexity of Definition~\ref{qc_perfect}.

The number of applications of classical Kolmogorov complexity
is countless, and it is our hope that this definition will
lead to a similar wide variety of applications in quantum 
complexity theory.

\section{Acknowledgements}  We would like to thank several
people for interesting discussions on this work:
Paul Vit\'anyi, Harry Buhrman, Richard Cleve, David Deutsch, Ronald de Wolf, 
John Watrous, Miklos Santha, Fr\'ed\'eric Magniez, and J\'er\'emy Barbay.

This work has been supported by Wolfson College Oxford,
Hewlett-Packard, European TMR Research Network ERP-4061PL95-1412,
the Institute for Logic, Language and Computation in Amsterdam,
an NSERC postdoctorate fellowship, and the EU fifth framework project 
QAIP IST-1999-11234.

\end{document}